\documentstyle[11pt,newpasp,twoside,epsf]{article}
\def\edcomment#1{\iffalse\marginpar{\raggedright\sl#1\/}\else\relax\fi}
\reversemarginpar

\newcommand{\bc}{\begin{center}}
\newcommand{\ec}{\end{center}}
\newcommand{\be}{\begin{equation}}
\newcommand{\ee}{\end{equation}}
\newcommand{\vs}{{\it vs.}}
\newcommand{\etal}{{\it et al.}}

\begin{document}
\title{Fitting Zodiacal Models}
 \author{Edward L. Wright}
\affil{Institute for Advanced Study, Princeton NJ 08540 and UCLA
Astronomy, PO Box 951562, Los Angeles CA 90095-1562}

\begin{abstract}
Models of the zodiacal light are necessary to convert measured data taken
from low Earth orbit into the radiation field outside the solar system.
The uncertainty in these models dominates the overall uncertainty in
determining the extragalactic background light for wavelengths 
$\lambda < 100\;\mu$m.
\end{abstract}

\section{Introduction}

The interplanetary dust particles responsible for the visible zodiacal light
absorb most of the solar radiation that falls on them, and reradiate it in
the infrared.  Thus a large part of the total infrared sky brightness from
space is due to the zodiacal dust cloud.  Modeling and removing this zodiacal
foreground is an important part of the analysis of data from any space-based
infrared experiment.  The {\sl IRAS} data have been fit both to physical
three-dimensional models of the zodiacal cloud (Good \etal, 1986) and to
phenomenological models of the variation with ecliptic latitude $\beta$
(Vrtilek \& Hauser, 1995).  The DIRBE data from {\sl COBE} have been fit to
3-D models by Kelsall \etal\ (1998) and Wright \etal\ (1998), and for all of
these fits the residuals are dominated by systematic errors in the 12 and
25 $\mu$m bands where the signal to noise ratio on the zodiacal emission is
high.

The fitting procedure used by Wright (1998) and Kelsall \etal\ (1998)
allows for an arbitrary galactic plus extragalactic signal at each pixel,
but this arbitrary flux must be constant in time.  All of the time variation
of the observed signal is assumed to be due to the changing line of sight
through the zodiacal cloud as the Earth orbits around the Sun.  Thus the model
fit to the data is
\be
I_{obs}(\lambda,l,b,t) = Z(\lambda,l,b,t;p) + I_c(\lambda,l,b)
\ee
The parameters of the model are the parameters $p$ of the zodiacal light model
plus the values $I_c$ -- one value for each band and observed spot on the sky.
There are thus a very large number of parameters in the model, but most of
them are in $I_c$ and can be found directly because they are {\em linear}
parameters.  There are 11 other linear parameters in the Wright models which
are the scattering efficiencies in bands 1-3 and the emission efficiencies
in bands 3-10.

I call this minimal assumption that the extrasolar system signal is
independent of time the ``weak no zodi'' principle.
For a spherical shell of dust at radius $R > 1$~AU, the RMS time variability
is only $\approx(4/R^2)\%$ of the total intensity, which is a very weak
signal indeed.

In order to save time, the model is only adjusted using a set of normal
points corresponding to a set of spots on the sky.  The main set of
points were selected by picking points with very small gradients in
$2^\circ$ patches both at 3.5 $\mu$m and 100 $\mu$m.  This selection reduces
the ``flicker noise'' associated with bright sources which are sometimes in and
sometimes out of the beam even while the beam center is always within a given
pixel; and it eliminates points near the galactic plane or bright stars or
bright cirrus clouds.
There are 399 of these patches.
An added set of points near the ecliptic but away from the galactic plane
are added when the parameters of the zodiacal bands are being determined.
Finally 1099 $5^\circ$ patches at $|b| > 20^\circ$ are used only for the 140
and 240 $\mu$m bands where the signal to noise ratio in the data is small.

\begin{figure}
\plotone{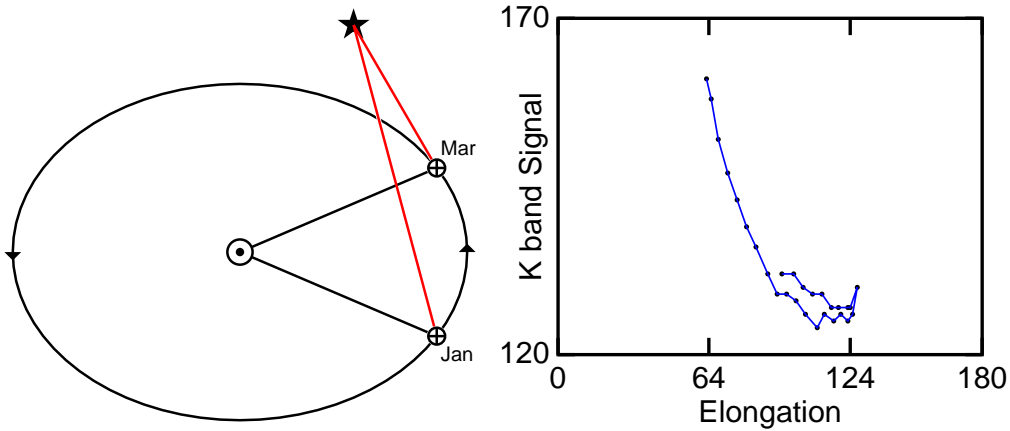}
\caption{Left: Looking at the same position at different times of year
gives different signals because of the different lines of sight through the
interplanetary dust cloud.  Right: The actual 2.2 $\mu$m signal \vs\ angle
between the line of sight (to the Gorjian, Wright \& Chary (2000) ``dark
spot'') and the Sun.\label{fig:zoft}}
\end{figure}

In this paper I try more powerful assumptions about the
celestial signal, and see how much leverage these give in fixing the
zodiacal light.  The first step in making assumptions about the sky is to
assume that the 25 $\mu$m sky at high $|b|$ is {\em isotropic}.  This is a very
reasonable assumption given the ratio of zodiacal to extrasolar system signals
in this bands.  This is equivalent to reducing the 399 separate parameters
in $I_c(25\;\mu\mbox{m},l,b)$ to a single parameter $I_c(25\;\mu\mbox{m})$.
I call this the ``strong no zodi'' principle.

The next step in making assumptions about the sky is to assume that the
high $|b|$ flux is {\em zero}.
Since the zodiacal emission is 10's of MJy/sr while
likely extragalactic background are only 10's of kJy/sr, this assumption may be
useful, but it is definitely a dangerous assumption to make when trying to find
an isotropic extragalactic background.  However, the fact that the model uses
independent emissivities in each band instead of assuming a smooth emissivity
\vs\ wavelength law means that the zodiacal model in the windows at 3.5 and
240 $\mu$m is somewhat isolated from the assumptions made at 25 $\mu$m.
Forcing the 25 $\mu$m fit to a zero extrasolar system signal changes the {\em
shape} of the cloud, and then this different shape leads to a different
magnitude of the 3.5 or 240 $\mu$m zodiacal flux.  I call this assumption
the ``very strong no zodi'' principle.

Two models have been converged to the DIRBE Pass 3B data using all
three fitting approaches.  One is a ``Good'' model of the type used by
Good \etal\ (1986).  The other model is ``FIZZ'', a fairly elaborate
physical model described in Wright (1998).

\bc
{\scriptsize
\quad\quad Table 1: Intensities in the Lockman Hole\\
\begin{tabular}{rrrrrrrrr}
\multicolumn{8}{c}{Intensity in MJy/sr in the Lockman Hole}\\
\hline
$\lambda\;[\mu$m] &
         GOOD1 &   GOOD2 &   GOOD3 &   FIZZ1 &   FIZZ2 &   FIZZ3 &  FIZZ3P & REALITY \\

1.25 &  0.1364 &  0.1407 &  0.1501 &  0.1696 &  0.1694 &  0.1806 &  0.1797 &  0.2228 \\
2.2  &  0.0942 &  0.0971 &  0.1037 &  0.1171 &  0.1170 &  0.1247 &  0.1240 &  0.1492 \\
3.5  &  0.0770 &  0.0788 &  0.0856 &  0.0897 &  0.0913 &  0.0976 &  0.0974 &  0.1149 \\
5    &  0.4287 &  0.4491 &  0.4699 &  0.4993 &  0.4954 &  0.5097 &  0.5088 &  0.5389 \\
12   & 13.4690 & 14.1695 & 14.7320 & 16.5817 & 16.2459 & 16.8580 & 16.8374 & 16.5239 \\
25   & 24.6018 & 27.2783 & 30.4484 & 29.2170 & 28.4249 & 30.2288 & 30.2234 & 30.0306 \\
60   &  6.8324 &  7.2153 &  7.4800 &  8.3259 &  8.0314 &  8.6145 &  8.6166 &  8.7382 \\
100  &  2.4155 &  2.5468 &  2.6346 &  3.0647 &  2.9380 &  3.1912 &  3.1932 &  4.1884 \\
140  &  1.1585 &  1.2219 &  1.3091 &  1.4817 &  1.4246 &  1.5838 &  1.5848 &  2.4480 \\
240  &  0.3571 &  0.3769 &  0.4028 &  0.4622 &  0.4414 &  0.4923 &  0.4927 &  0.9459 \\
\hline
\end{tabular}
}
\ec

The values of the zodiacal models, averaged over the actual
observation times, for a set of pixels in the Lockman hole, are
given in Table 1.
GOOD1 and FIZZ1 use the ``weak no zodi principle'', GOOD2 and FIZZ2 the
``strong no zodi principle'', and GOOD3 and FIZZ3 the ``very strong no zodi
principle''.
REALITY in Table 1 is the average of the actual DIRBE data.
The ratio of 240 to 25 $\mu$m flux is fairly constant for
the FIZZ models, ranging from 0.0155 to 0.0163.  This suggests that a simple
calculation that assigns an uncertainty to the 25 $\mu$m background can be used
to determine the uncertainty of the zodiacal light model in the 240 $\mu$m
band.  Thus simple fitting procedures such as the one used by
Schlegel, Finkbeiner \& Davis (1998) should be adequate for determining the 240
$\mu$m background.  At 3.5 $\mu$m the situation is more ambiguous because of
the scattered radiation which amounts to about 50\% of the total flux in this
window.  Changes in the assumed phase function 
$\Phi$ for scattering have no effect
whatsoever at 25 $\mu$m but can have significant effects at 3.5 $\mu$m.
Thus the model changes produced by the three different fitting procedures
considered here do not adequately span the total range of possible models
and could underestimate the systematic errors in the zodiacal flux in the short
wavelength window.
Even so, letting $\ln\Phi$ be a quartic (FIZZ3P in Table 1)
instead of quadratic polynomial
in $\mu$, the cosine of the scattering angle, produces only a 0.2~kJy/sr
change in the zodiacal light model at 3.5 $\mu$m in the Lockman hole.

\end{document}